\documentclass[runningheads]{llncs}
\usepackage{graphicx}
\usepackage{float}
\usepackage{BOONDOX-calo}
\usepackage{multirow}
\usepackage{multicol}
\usepackage{url}
\usepackage{graphicx}
\graphicspath{ {figures/} }
\usepackage{amsxtra}
\usepackage{fancyhdr}
\usepackage{supertabular}
\usepackage{amssymb}
\usepackage{amsmath}

\usepackage{epsfig}
\usepackage[english]{babel}
\usepackage{graphics}
\usepackage{setspace}
\usepackage{epsfig}
\usepackage{amssymb}
\usepackage{amsmath}
\usepackage{cite}
\usepackage{graphicx}
\usepackage{listing}
\usepackage{times}
\usepackage{fancyhdr
}
\usepackage{supertabular}
\usepackage[utf8]{inputenc}
\usepackage{dirtytalk}
\usepackage{amssymb}
\usepackage{amsmath}
\usepackage{epsfig}
\usepackage{verbatim}
\usepackage{setspace}

\usepackage[linesnumbered,ruled,vlined]{algorithm2e}
\SetKwInput{KwInput}{Input} 
\SetKwInput{KwOutput}{Output}
\usepackage{tocloft}
\usepackage{graphicx}
\usepackage{enumerate}
\usepackage[usenames]{color}
\usepackage{amsfonts}
\newcommand{\R}{\mathbb{R}}

\begin{document}
\title{Effect of Perturbation and Topological Structure on Synchronization Dynamics in Multilayer Networks}
%
%
\author{Rajesh Kumar \inst{1} \and
Suchi Kumari \inst{2} \and Anubhav Mishra \inst{3}}
\authorrunning{Kumar et al.}
%
\institute{Amity School of Engineering and Technology, Amity University, Bangalore, India. \email{rkumar1@blr.amity.edu} \and
Department of Computer Science Engineering, Shiv Nadar Institute of Eminence, Delhi-NCR, India.
\email{suchi.singh24@gmail.com} \and
Ghani Khan Choudhary Institute of Engineering and Technology, Malda, West Bengal, India
\email{tooli.anubhav@gmail.com}}
\maketitle              
\begin{abstract}
The way the topological structure transforms from a decoupled to a coupled state in multiplex networks has been extensively studied through both analytical and numerical approaches, often utilizing models of artificial networks.  These studies typically assume uniform interconnections between layers to simplify the analytical treatment of structural properties in multiplex networks. However, this assumption is not applicable for real networks, where the heterogeneity of link weights is an intrinsic characteristic. Therefore, in this paper, link weights are calculated considering the node’s reputation and the impact of the inter-layer link weights are assessed on the overall network's structural characteristics. These characteristics include synchronization time, stability of synchronization, and the second-smallest eigenvalue of the Laplacian matrix (algebraic connectivity). Our findings reveal that the perturbation in link weights (intra-layer) causes a transition in the algebraic connectivity whereas variation in inter-layer link weights has a significant impact on the synchronization stability and synchronization time in the multiplex networks. This analysis is different from the predictions made under the assumption of equal inter-layer link weights.

\keywords{Multiplex network  \and Algebraic connectivity \and Perturbation \and Synchronization  \and Time scale \and Stability.}
\end{abstract}
\section{Introduction}
One fundamental property of connected identical dynamical systems is synchronization. Within multiplex networks, synchronization dynamics span multiple layers and occur within individual network layers. The algebraic connectivity, commonly represented by the second smallest eigenvalue ($\lambda_2$) of the multiplex network's Supra-Laplacian matrix, is closely associated with these synchronization patterns. Research on synchronization dynamics is crucial for various real-world systems in information technology, engineering, biology, physics, and social science \cite{strogatz2000kuramoto}. Therefore, in this research, we examined how the algebraic connectivity ($\lambda_2$) in multiplex networks is influenced by the topological structure. We then demonstrated the stability of synchronization processes in weighted multiplex networks, followed by an analysis of synchronization time.

The spectral characteristics of the Supra-Laplacian matrix in multiplex networks are crucial for understanding synchronization dynamics. Unlike the adjacency matrix, the eigenvalues of the Laplacian matrix provide clearer insights \cite{mohar1991}. Specifically, algebraic connectivity, indicated by the second smallest eigenvalue ($\lambda_2$), offers valuable information about the network's modularity \cite{newman2006} and synchronization capacity \cite{aguirre2014}. Synchronization dynamics in multiplex networks are influenced by variables such as average node degree, clustering coefficients in each layer, and inter-layer link weights. Erratic shifts in algebraic connectivity often signify faster diffusion due to additional paths between nodes in the multilayer structure \cite{de2016,kumar2021minimizing}. For single-layer networks, synchronization stability is quantified using the eigenratio $R=\lambda_{N}/\lambda_{2}$, where $\lambda_2$ and $\lambda_{N}$ are the second smallest and largest eigenvalues of the Laplacian matrix \cite{barahona2002}. Radicchi et al. \cite{radicchi2013} explored how changes in the second smallest eigenvalue of the supra-Laplacian matrix lead to the emergence of two distinct regimes and a structural transition phase. This observation highlights the significant influence of structural factors on network dynamics.

The effect of inter-layer link patterns on the dynamics of spreading processes in an interconnected network has been studied by researchers in some recent work. Wang et al. \cite{wang2011} demonstrated that inter-layer connections based on node degrees have a relatively smaller effect on the size of infection compared to the density of interconnections. In another study, the authors investigated the correlation between intra-layer and inter-layer degrees \cite{saumell}. They found that a strong correlation between these degrees can lead to the outbreak state, even if the epidemic threshold is not reached. When dealing with multiplex networks, the influence of their topological properties can become more complex. Simply examining a network layer in isolation, without considering its interactions with other layers, can lead to misleading conclusions. 

Aguirre et al. \cite{aguirre2014synchronization} investigated the impact of the connector node degree on the synchronizability of two-star networks with one inter-layer link. They showed that synchronization could be achieved by connecting the high-degree nodes of each network. Xu et al. \cite{xu2015synchronizability} studied the synchronizability of two-layer multiplex networks for three different coupling patterns. They determined that there exists an optimal value of the inter-layer coupling strength for maximizing complete synchronization. Authors in \cite{serrano2017optimizing} demonstrated the impact of topological similarity among multiplex layers on synchronization performance. The distance between the pairs of layers is used to measure the similarity between the network layers \cite{andrade2008measuring}. They fixed one network layer while rewiring the network layers. The authors discovered that a sizeable inter-layer coupling generally promotes global multiplex synchronization for the fixed intra-layer and inter-layer diffusion coefficients.

In networks like Small World (SW) and Scale-Free (SF) characteristics, larger values of the average clustering coefficient ($\langle CC \rangle$) can hinder global synchronization. This occurs because the network tends to divide into clusters that oscillate at different frequencies \cite{mcgraw2005}. Let us take the example of interconnected Facebook and Twitter social networks. When viewing both networks as interconnected multiplex networks, it has an impact on their structural characteristics. Assume that a node $i$ (part of the connected triangle) has a clustering coefficient value of  $1$ in a single independent network layer.  However, when node $i$ is connected to another node, say $j$, in a different network layer, it alters the clustering coefficient of node $i$. This demonstrates that the topological structures of multiplex networks differ from those of their individual network layers. In the above mentioned research work, in impact of ($\langle CC \rangle$) on the synchronization is studied on the independent networks. However, real-world systems consist multiple subsystems which are inter-connected and inter-dependent e.g., physical system, infra-structure systems, biological system etc. Therefore, in the presented research work, effect of average clustering coefficient ($\langle CC \rangle$) on the stability of the synchronization in the multiplex networks is analyzed. Besides, edge weight in the networks is an abstraction and can be assigned in multiple ways. However, in real world networks, strength of the relationship is function of nodes connecting the edges. For example, number of interaction may be accounted as edge weights. In this research work, a method is proposed to calculated the edge weights which is function of the trust relationship among the pair of nodes. Hence, depending upon the network structure of individual layers, the behavior of $R$ in the multiplex networks changes with the variation in inter-layer link weights \cite{sole2013}. In certain cases, a decrease in the value of eigenratio $R$ = ($\lambda_{N}/\lambda_2$) indicates that the two layers of the multiplex network begin to synchronize and behave as if they were a single-layer network \cite{sole2013}. Their investigations showed that for various topologies, for weaker inter-layer coupling weight, the value of parameter $R$ of the multiplex network is approximately the same as the value of $R$ of the individual network layer having the highest value of $\lambda_2$ of the Laplacian matrix. For decisive inter-layer coupling weights parameter, $R$  for the multiplex-network is approximated by taking the ratio of $\lambda_{N}$ and $\lambda_2$ of the average Laplacian matrix of the layers of the multiplex-network. These findings are a direct consequence of the strong inter-layer coupling strength between two layers\cite{sole2013}. In the previous research works, the behavior of stability has been studied by considering the unweighted and undirected multiplex networks \cite{gomez,sole2013}. However, real multiplex networks \cite{kumar2023robustness} are often weighted and exhibit distinct topological properties. Hence, in this paper, we address this gap to study the synchronization dynamics in weighted multiplex networks of different topological structures by introducing the perturbation to the inter-layer edge weights. Also,  Algebraic connectivity \cite{de2007} is closely related to synchronization processes in multiplex networks and is used to study diffusion characteristics \cite{gomez}, synchronization ability \cite{aguirre2014}, and modularity \cite{newman2006}. The spectral characteristics of the combinatorial Supra-Laplacian matrix significantly influence the dynamics of these synchronization processes. Therefore, impact of perturbation on the Algebraic connectivity \cite{de2007} is analyzed by considering the different topological structures of synthetic multiplex (Random and Power law) networks and real world multiplex networks \cite{magnani2013}. 
 
Major contributions of the presented research work are as follows: 
\begin{itemize}
   \item {A method is proposed to compute the intra-layer and inter-layer link weights.} 
   \item{Investigate the effect of perturbation (in the network layers) on the algebraic connectivity ($\lambda_2$) for multiplex networks with different topological structures.} 
   \item{Analyze the stability of synchronization dynamics on considered multiplex networks by varying the inter-layer link weights.}
   \item{Study the time scale (level of synchronization) for the multiplex networks.} 
   \item{Compare the findings of unweighted and weighted synthetic multiplex networks as well as empirical dataset multiplex networks} 
\end{itemize}

The remainder of the paper is organized as follows: Section \ref{proposed_method} provides information on the proposed methods for computing weights. Section \ref{simulation_setup} describes the details of the simulation setup used in our study. The analysis and results are presented in Section \ref{result_analysis_ch3}. Finally, Section \ref{conclusion_ch3} concludes with the research findings and suggests possible directions for future investigation.

\section{Proposed Methodology} \label{proposed_method}
\begin{figure}[!htb]
\begin{center}
\includegraphics[width=0.8\linewidth,height=2in]{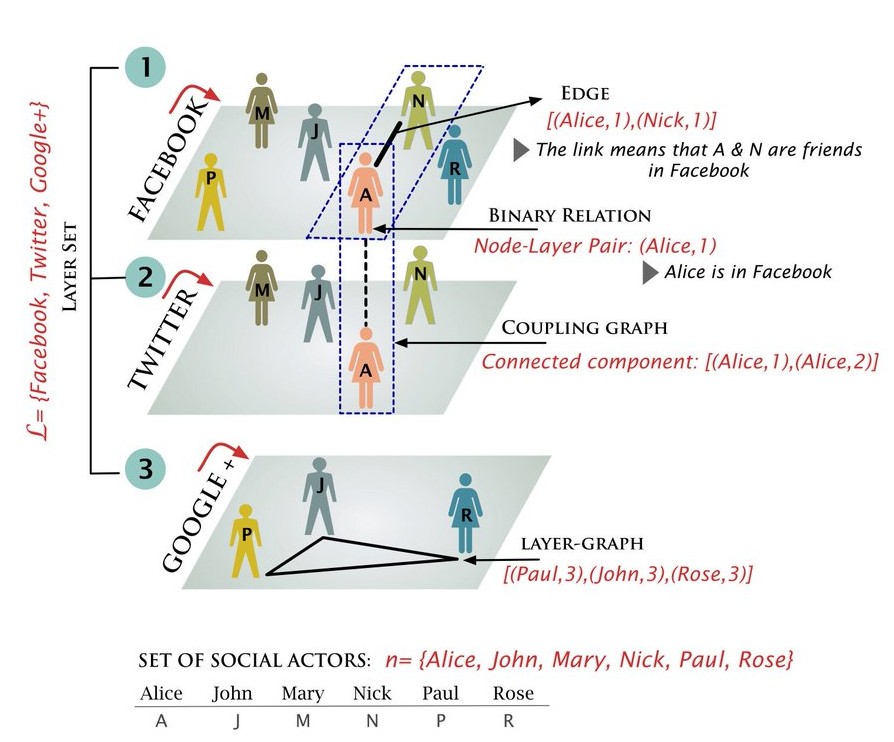} 
\caption{Schematic representation of multiplex network composed by online social
networks \cite{de2018fundamentals}}

\label{ch3_mln_example}
\end{center}
\end{figure}

A multiplex network is represented as a pair $\mathcal{M} = (\mathcal{G},\mathcal{E})$, where $\mathcal{G}$ = $\{G_{\alpha}; \alpha\in $ $\{1,2,\dots\ ,M \}$   $\}$ is a set of $M$ directed/undirected, weighted/unweighted graphs $G_\alpha=(\mathcal{V}_\alpha,\mathcal{E}_\alpha)$ (called network layers of $\mathcal{M}$) \cite{kivela2014multilayer}. Each layer of Multiplex networks contains same set of nodes ($\mathcal{V}_\alpha=\mathcal{V}_\beta=\mathcal{V}$) and each node in network layer $G_{\alpha}$ is connected to its replica node in network layer $G_{\beta}$ i.e., $\mathcal{E}_{\alpha \beta}=\{(i^{\alpha},i^{\beta}); i^{\alpha}\in \mathcal{V}_{\alpha}, i^{\beta}\in \mathcal{V}_{\beta}$ for every  $1 \leq \alpha \neq \beta \leq M$. A good example is social multiplex networks where a group of users are present in each network layer (Facebook, Twitter etc.) shown in Fig. \ref{ch3_mln_example}, but these users may have different friends in each network layer. 

In matrix notation, multiplex networks are represented as supra-adjacency matrix $\mathcal{A}^{\mathcal{M}}$.  
  
\[
\mathcal{A}^{\mathcal{M}}=
\left[
\begin{array}{c|c}
\mathcal{A}^{[\alpha]} & \mathcal{A}^{[\alpha \beta]} \\
\hline
\mathcal{A}^{[\alpha \beta]} & \mathcal{A}^{[\beta]}

\end{array}
\right]
\]
where $\mathcal{A}^{[\alpha]}=a_{ij}^{\alpha} \in \R^{N_{\alpha}\times N_{\alpha}}$ is adjacency matrix of network layer $G_\alpha$  \cite{boccaletti2014structure} with element,

\[
    a_{ij}^{\alpha}=\left\{
                \begin{array}{ll}
                  1 \hspace{2mm}if\hspace{1mm} (e_{i,j}^{\alpha})\in {E}_{\alpha}  \\
                 
                  0 \hspace{2mm} otherwise
                \end{array}
              \right\}
  \]
and $N_{\alpha}$ is the number of nodes in the network layer $G_\alpha$. The inter-layer adjacency matrix $\mathcal{A}^{[\alpha \beta]}$ with element $(a_{ij}^{\alpha \beta}) \in \R^{N_{\alpha} \times N_{\beta}}$ \cite{boccaletti2014structure} such that,

  \[
    a_{ij}^{\alpha \beta}=\left\{
                \begin{array}{ll}
                  1 \hspace{2mm}if\hspace{1mm} (e_{i,j}^{\alpha,{\beta}})\in {E}_{\alpha {\beta}}  \\
                 
                  0 \hspace{2mm} otherwise
                \end{array}
              \right\}
  \]
In this section, we first describe the method to compute link weights, and then the stability of synchronization and synchronization time of the entire multiplex network is discussed.

\subsection{Intra-layer and inter-layer Edge Weights Calculation}

Network edge weights can be conceptually interpreted in various ways depending on the which type pf network we are considering. For example, in social networks, tracking individual interactions over a specified period might provide a representative weight or indicator of link strength. In such networks, the edge weights can be calculated by combining the node's trust value and the degree of connectivity between nodes in the network. The node's trust score can be evaluated using the well-known statistical method, Maximum Likelihood Estimation. Th evaluate the overall trust value ($\phi^{\mathcal{M}}_{i}$) associated with a particular node $i$, we define matrices $\mathcal{S}^{\mathcal{M}}$ and $\mathcal{F}^{\mathcal{M}}$, of dimensions ($M \times N$), where $M$ is the number of layers in the multi-layer network and $N$ is the total number of nodes. These matrices record the number of ``successful" and "failed" transactions between nodes $i$ and $j$ within the multi-layer network. The formulation is given as follows:

\[
    s_{ij}= 
\begin{cases}
    T^{\alpha}_{ij} +  T^{\alpha \beta}_{ij} ,& \text{if there is an edge between} \hspace{2mm}i^{\alpha} \text{and} j^{\alpha},  i^{\alpha} \text{and} j^{\beta}: \alpha \neq \beta\\
    0,              & \text{otherwise} 
\end{cases}
\]

\[
    f_{ij}= 
\begin{cases}
    U^{\alpha}_{ij} +  U^{\alpha \beta}_{ij} ,& \text{if there is an edge between} \hspace{2mm}i^{\alpha} \text{and} j^{\alpha},  i^{\alpha} \text{and} j^{\beta}: \alpha \neq \beta\\
    0,              & \text{otherwise} 
\end{cases}
\]
where  ($s_{ij} \in \mathcal{S}^{\mathcal{M}}$,  $f_{ij} \in \mathcal{F}^{\mathcal{M}}$) are the counts of successful and failed transactions between nodes $i$ and $j$. The terms $\psi^{\mathcal{M}}_{i}=\displaystyle \sum_{i=1,i \neq j}^{M\times N}s_{ij}$ and $\theta^{\mathcal{M}}_{i}=\displaystyle \sum_{i=1,i \neq j}^{M\times N}f_{ij}$ hold the aggregate of successful and failed transaction so that likelihood function of ($\phi^{\mathcal{M}}_{i}$) is given in Eq. \eqref{eq1}.
\begin{equation}
    L(\phi^{\mathcal{M}}_{i},y_{i,1},y_{i,2},\dots,y_{i,MN})=\displaystyle \prod_{j=1,j \neq i}^{MN} ({\phi^{\mathcal{M}}_{i}})^{\psi^{\mathcal{M}}_{i}} (1-\phi^{\mathcal{M}}_{i})^{\theta^{\mathcal{M}}_{i}}
    \label{eq1}
\end{equation}
Using Eq. \eqref{eq1}, the Log likelihood function ($\zeta$) of $\phi^{\mathcal{M}}_{i}$ can be recasted as,
\begin{equation}
    \zeta(\phi^{\mathcal{M}}_{i})=\text{ln} \hspace{1mm}L(\phi^{\mathcal{M}}_{i})=\psi^{\mathcal{M}}_{i} \text{ln}\hspace{1mm}\phi^{\mathcal{M}}_{i} + \theta^{\mathcal{M}}_{i} \text{ln}\hspace{1mm} (1-\phi^{\mathcal{M}}_{i})
    \label{eq2}
\end{equation}
Likelihood expression is obtained by taking derivative of the Eq. \eqref{eq2} and represented in Eq. \eqref{eq3}
\begin{equation}
    \frac{d\zeta(\phi^{\mathcal{M}}_{i})}{d\phi^{\mathcal{M}}_{i}}=\frac{\psi^{\mathcal{M}}_{i}}{\phi^{\mathcal{M}}_{i}}-\frac{\theta^{\mathcal{M}}_{i}}{1-\phi^{\mathcal{M}}_{i}}=0
    \label{eq3}
\end{equation}
The derivative in Eq. \eqref{eq3} provides the the trust score of node $i^{\mathcal{M}}$ using Maximum Likelihood Estimation of $(\phi^{\mathcal{M}}_{i}$) to maintain quality of services (QoS). After evaluating the trust score of each node in the multiplex network, the edge weight can be evaluated in inter-layer as well as in intra-layer.
 
 \subsubsection{Computation of intra-layer and inter-layer edge weights: } \label{edge_weights}
 
Connection patterns ($a_{ij}^{\alpha}$) and trust values ($\phi^{\alpha})$ of the nodes are taken into account while calculating intra-layer and inter-layer link weights ($w_{ij}$). The link weights are calculated using $\Gamma$ function in Eq. \eqref{eq5}. 

\begin{equation}
    w_{ij}^{\alpha}=\mathcal{F}(\Gamma,a_{ij}^{\alpha})=\Gamma+a_{ij}^{\alpha}
    \end{equation}
    \begin{equation}
        \Gamma(\phi_{i}^{\alpha},\phi_{j}^{\alpha})=\Big{[\Big(} \frac{\text{Cos}(\Delta_{ij}^{\alpha})+1}{2}\Big{)}\times(\phi_{i}^{\alpha}\times\phi_{j}^{\alpha})\Big{]} \label{eq5}
    \end{equation}

The $\Delta$ function is used to identify like-minded nodes in the network. In Eq. \eqref{eq5}, $\Delta_{ij}^{\alpha}$ is evaluated by taking the difference ($|\phi_{i}^{\alpha} - \phi_{j}^{\alpha}|$) of the trust scores of nodes $i$ and $j$. Since the trust score is a probability function, its value will lie in the range $[0,1]$. The $\cos(\Delta_{ij})$ value, however, will be any real number in the range $[-1,1]$. To normalize the $\cos(\Delta_{ij})$ score within the range $[0,1]$, it is transformed to $\frac{\cos(\Delta_{ij}^{\alpha}) + 1}{2}$.

Let us examine two different situations:
\begin{enumerate}[(i)]
   \item  $\phi_{i}^{\alpha}=0.1$ and $\phi_{j}^{\alpha}=0.1$.
    \item $\phi_{i}^{\alpha}=0.99$ and $\phi_{j}^{\alpha}=0.99$.
\end{enumerate}

Since the computed trust difference in both scenarios is $\Delta_{ij}^{\alpha}=0$, the value of $\frac{\cos(\omega)+1}{2}$ is $1$. Although both scenarios provide the same score, the trust scores are noticeably greater in the second scenario compared to the first. To rectify this disparity, a scaling factor is implemented, and the trust values are adjusted accordingly. The trust values for nodes $i$ and $j$ are multiplied ($\phi_{i}^{\alpha} \times \phi_{j}^{\alpha}$) by the normalized cosine function $(\frac{\cos(\omega)+1}{2})$ to ensure the trust values are adjusted in a way that appropriately represents their magnitudes.

\subsection{Stability of Synchronization} \label{ch3_stability}
Suppose that each node $i$ at any layer $L^{\alpha}$ has some amount of information $x_i^{\alpha}(t)$ at some time $t$. Assume that this information flows from node $j^{\alpha}$ to node $i^{\alpha}$ at rate $p(x_j^{\alpha} - x_i^{\alpha})$ in network layer $L^{\alpha}$ and from node $i^{\alpha} \in L^{\alpha}$ to node $i^{\beta} \in L^{\beta}$ with rate $d_x(x_i^{\beta} - x_i^{\alpha})$, where $p$ and $d_x$ are the parameters controlling the intra-layer and inter-layer weights respectively. The dynamical equation \cite{gomez} characterizing the evolution of the state of node $x_i^{\alpha}$ at network layer $L^{\alpha}$ in a multiplex network having $M$ networked layers is given in Eq. \eqref{ch3_dx}.

\begin{equation}
\frac{dx_{i}^{\alpha}}{dt}=p \sum_{j=1}^{N}w_{ij}^{\alpha}(x_{j}^{\alpha}(t)-x_{i}^{\alpha}(t)) + d_x\sum\limits_{\substack{\beta=1 \\ \alpha \neq \beta}}^{M}w_{ii}^{\alpha\beta}(x_{i}^{\beta}(t)-x_{i}^{\alpha}(t)) \hspace{5mm}
\label{ch3_dx}
\end{equation}
where $w_{ij}^{\alpha}$ and $w_{ii}^{\alpha \beta}$ are the edge weights of intra-links and inter-links, respectively. The parameters $p$ and $d_x$ are considered as tuning parameters to control $w_{ij}^{\alpha}$ and  $w_{ij}^{\alpha \beta}$, respectively. The rate equation in Eq. \eqref{ch3_dx} can be represented in the matrix form in Eq. \eqref{ch3_x}.

\begin{equation}
\dot{\textbf{x}} = (p \mathcal{L}^L + d_x \mathcal{L^I})\textbf{x} = \mathcal{L^{M}}\textbf{x}
\label{ch3_x}
\end{equation}

where $\mathcal{L}^L$ and $ \mathcal{L^I}$ depicts the supra-Laplacians of the intra-layer and inter-layer networks, respectively and $d_x$ controls the inter-layer link weights \cite{gomez}. The Laplacian matrix of each network layer $\mathcal{L}^\alpha$ is formulated as $\mathcal{L}^\alpha$ = $\Delta^\alpha$ - $W^\alpha$ where $W^\alpha$ is the weight matrix of network layer $L^{\alpha}$ with elements $w_{ij}^{\alpha}>0$ if there is an edge between $i^{\alpha}$ and $j^{\alpha}$, otherwise $w_{ij}^{\alpha}=0$. The matrix ${(\Delta^\alpha)}$ is diagonal matrix with elements $\delta_{ii} = \sum_{j}w_{ij}^\alpha$. Considering both the parameters ${\Delta^\alpha}$ and $W^\alpha$, the matrix $\mathcal{L}^L$ is provided in Eq. \eqref{ch3_lap}.

\begin{equation}
\mathcal{L}^L=\begin{pmatrix}
p^1 \mathcal{L}^{(1)} & 0 & \cdots &  0 \\
 0   &   p^2 \mathcal{L}^{(2)} &\cdots & 0 \\
 \vdots & \vdots &\ddots & \vdots  \\
 0 & 0 & \cdots &  p^M\mathcal{L}^{(M)} \
\end{pmatrix} \vspace{3mm}
\label{ch3_lap}
\end{equation}

and inter-layer weight matrix is given in Eq. \eqref{ch3_ilw}

\begin{equation}
\mathcal{W^I}=\begin{pmatrix}
0 & \mathcal{W}_{(1,2)}I & \mathcal{W}_{(1,3)}I&\cdots &  \mathcal{W}_{(1,M)}I \\
 \mathcal{W}_{(2,1)}I   & 0 & \mathcal{W}_{(2,3)}I   &\cdots & \mathcal{W}_{(2,M)}I \\
 \mathcal{W}_{(3,1)}I   & \mathcal{W}_{(3,2)}I & 0   &\cdots & \mathcal{W}_{(3,M)}I \\
 
 \vdots & \vdots && \ddots & \vdots \\
 \mathcal{W}_{(M,1)}I & X_{(M,2)}I& X_{(M,3)}I & \cdots &  0 \
\end{pmatrix}
\label{ch3_ilw}
\end{equation}

where $\mathcal{W}_{(\alpha \beta)}$ is a vector of inter-layer edge weights \{ $w_{ii}^{\alpha\beta}$\} connecting $i^{\alpha}\in L^\alpha$ and $i^{\beta} \in L^{\beta}$, $I$ is the identity matrix. Inter-layer Laplacian $\mathcal{L^I}$ matrix can be determined from $\mathcal{W}^I$. The state of a node $i$ in Eq. \ref{ch3_dx} can be found using $x_i(t) = x(0)e^{-\lambda_i t}$ \cite{newman2018networks,gomez2013diffusion}, where $\lambda_i$ is the eigenvalue of supra-Laplacian matrix $\mathcal{L^{L}}$. Apart from that, the second smallest eigenvalue ($\lambda_2$) of the Laplacian matrix is considered as algebraic connectivity of the graph $G$ because of the following proposition:\\

\textbf{Proposition}: \textit{Let $G$=$(V,E)$ be a connected graph  with positive weights $w_{ij}$, then the algebraic connectivity of graph $G$ is positive and equal to the minimum of the function \cite{de2007}} 
\begin{equation}
\zeta(x)=N\frac{\sum_{i,j\in E}w_{ij}(x_i-x_j)^2 }{\sum_{i,j\in E,i<j}w_{ij}(x_i-x_j)^2}
\label{ch3_algebraic}
\end{equation}
\textit{over all non-constant $N$-tuples $x=(x_i)$}.\\

Stability of the synchronization is determined by the parameter $R = \frac{\lambda_{N}}{\lambda_2}$ \cite{barahona2002}, where $\lambda_N$ and $\lambda_2$ are the largest and smallest non-zero eigenvalues of the matrix $\mathcal{L^{M}}$. For the lower values of $d_x$, $\lambda_2$ of $\mathcal{L^{M}}$ is approximated by $\lambda_2 (\mathcal{L}) \approx \lambda_2 (\mathcal{L^I})$. Thus the behavior of $R$ can be understood by the approximation \cite{sole2013},
\begin{equation}
R \approx \frac{max(\lambda_{N}(\mathcal{L}^{\alpha})+s_{i}^\mathcal{I})}{\lambda_2(\mathcal{L^I})d_x}
\label{ch3_weak}
\end{equation}
where $\mathcal{L}^\alpha$ is the Laplacian matrix of layer $L^{\alpha}$, $\lambda_{N}$ is the maximum eigenvalue of $\mathcal{L}^\alpha$ and $s_{i}^\mathcal{I}=\sum\limits_{\substack{\beta=1 \\ \alpha \neq \beta}}^{M}w_{ii}^{\alpha\beta}$ is the strength of the node $i$ in the inter-layer network having maximum eigenvalue \cite{sole2013}.\\
For the strong inter-layer link weights ($d_x >> 1$), $R$ is approximated as,
\begin{equation}
R\approx \frac{d_x\lambda_{M}(\mathcal{L^I})+\lambda_{N}(\mathcal{L}_{max}^{WA})}{\lambda_2(\mathcal{L}^{AV})}
\label{ch3_strong}
\end{equation}
where $\mathcal{L}^{AV}$ is the average Laplacian of network layers and $\mathcal{L}_{max}^{WA}$ is given by

\begin{equation}
\mathcal{L}_{max}^{WA} = \frac{1}{\parallel {\textbf{x}}^{'\mathcal{I}} \parallel^2}\sum_{\alpha}(x_{\alpha}^{'\mathcal{I}})\mathcal{L}^{\alpha}
\label{ch3_lmax}
\end{equation}
where ${\textbf{x}}^{'\mathcal{I}}$ is the eigenvector corresponding to maximum eigenvalue of inter-layer network matrix\cite{sole2013}. 

\subsection*{Synchronization Time} \label{ch3_sync_time}
Let us assume that each node of the network layer is embedded within an oscillator and the level of synchronization on the $MN$ interaction units can be expressed by quantity $S(\tau)$ for a large timestamps $\tau$ in Eq. \eqref{ch3_time}.

\begin{equation}
S (\tau)=\frac{1}{MN}\sum_{i=1}^{MN}(1-x_i(\tau))
\label{ch3_time}
\end{equation}
where $x_i(\tau) = x_i(0)e^{-\lambda_i \tau}$ and $\lambda_{i}$ is the $i^{th}$ eigenvalue of the supra-Laplacian matrix. The value of synchronization score $S \rightarrow 1$, when the system is completely synchronized and for sufficiently large timestamps $\tau$. The corresponding time is known as synchronization time $\tau_s$, starting from random phase $x_i$ \cite{guardiola2000} of node $i$ in any network layer.

\section{Simulation Setup} \label{simulation_setup}
For the simulation, two versions of synthetic multiplex networks with two layers are considered. The first is constructed using the Barabasi-Albert model of preferential attachment \cite{albert2002}, with network layers $L_1$ and $L_2$. Each layer contains 200 nodes, with $m = 2$ for layer $L_1$ and $m = 3$ for layer $L_2$, where $m$ is the number of stubs with which a new node attaches to existing nodes in the network layers. The second network is designed using the Power law network model ($p(k) \sim k^{-\gamma}$), with $\gamma = 2.1$ for layer $L_1$ and $\gamma = 2.2$ for layer $L_2$. Each layer in this network also contains 200 nodes. After constructing the network layers, inter-layer connections are established so that each node in layer $L_1$ is connected to its replica node in layer $L_2$, as shown in Fig. \ref{ch3_mln_example}. For validation, a multiplex network consisting of two layers is designed using the empirical dataset from the CS-Aarhus social multiplex network \cite{magnani2013}. The parameters related to the dataset are shown in Table \ref{ch3_real_tab}. Two network layers, Facebook and Lunch, are considered for analysis due to their single connected component.

\begin{table}[htb!]
\begin{center}
\caption{ Network layer parameters of the dataset CS-Aarhus social multiplex network \cite{magnani2013}}
\begin{tabular}{|p{2.5cm}|p{1.5cm}|p{1.7cm}|p{1.9cm}|p{1.6cm}|p{1.75cm}|}
\hline
Name of network layer & Nodes  & Edges & Conn. Comp. &  Average node degree  \\

\hline
 Work & 61 & 194 & 2 & 6.47 \\
\hline
Leisure & 61 & 88 & 1 & 3.74 \\
\hline
 Coauthor & 61 & 21 & 8 & 1.68\\
\hline
Lunch & 61 & 193 & 1 & 6.43 \\
\hline
Facebook & 61 & 124 & 1 & 7.75\\
\hline
\end{tabular}
\label{ch3_real_tab}
\end{center}
\end{table}

To analyze the effect of perturbation on the inter-layer links in the behavior of algebraic connectivity ($\lambda_2$), stability of the synchronization and synchronization time $(\tau)$, following scenario is considered for unweighted and weighted multiplex networks.

\begin{itemize}

\item{For weighted multiplex networks, intra-layer edge weights $w_{ij}^\alpha$ and inter-layer edge weights $w_{ii}^{\alpha \beta}$ are computed according to the method discussed in Section \ref{edge_weights}}
\item{For unweighted case, values of $w_{ij}^\alpha$ and $w_{ii}^{\alpha \beta}$ are taken as $1$.}
\item{For weighted and unweighted multiplex networks, value of $d_x$ and $p$  is taken from $0$ to $2$ in the step-size of $0.001$.}
\end{itemize}

\section{Result and Analysis} \label{result_analysis_ch3}
In this section, we analyze the effect of perturbations on algebraic connectivity, synchronization stability, and synchronization time in the multiplex network.

\subsection{Algebraic Connectivity} \label{ch3_result_algebraic}

We study the behaviour of algebraic connectivity ($\lambda_2$) for the considered synthetic as well empirical data set multiplex network (weighted and unweighted). For $d_x$ = 0, spectrum of the supra-Laplacian matrix of the multiplex network is given by $\wedge$($\mathcal{{L}^M}$) = $\{0=\lambda_1= \lambda_2<\lambda_3\leq\lambda_4\dots\lambda_{2N} \}$ with $\lambda_3$ = \textit{min}($\lambda_{2}^1$,$\lambda_{2}^2$) where $\lambda_{2}^1$ and $\lambda_{2}^2$ are the second smallest eigen values of Laplacian matrices of first and second network layers of multiplex networks for unweighted and weighted cases \cite{gomez2013diffusion}. The spectrum of the inter-layer Laplacian matrix is given by $\wedge(\mathcal{L}^I) = {0, 2 d_x}$. In the absence of inter-layer edges, the two network layers are isolated, resulting in $\lambda_1$ and $\lambda_2$ both being 0 in the spectrum $\wedge(\mathcal{L}^M)$. To study the effect of inter-layer edge weights on the evolution of algebraic connectivity, we consider the following synthetic multiplex networks. 

\subsubsection{Network Layers of Multiplex Network Designed Using BA Model} \label{ch3_result_algebraic_BA}

\begin{figure}[!h]
\begin{center}
\includegraphics[width=1.0\linewidth,height=2.in]{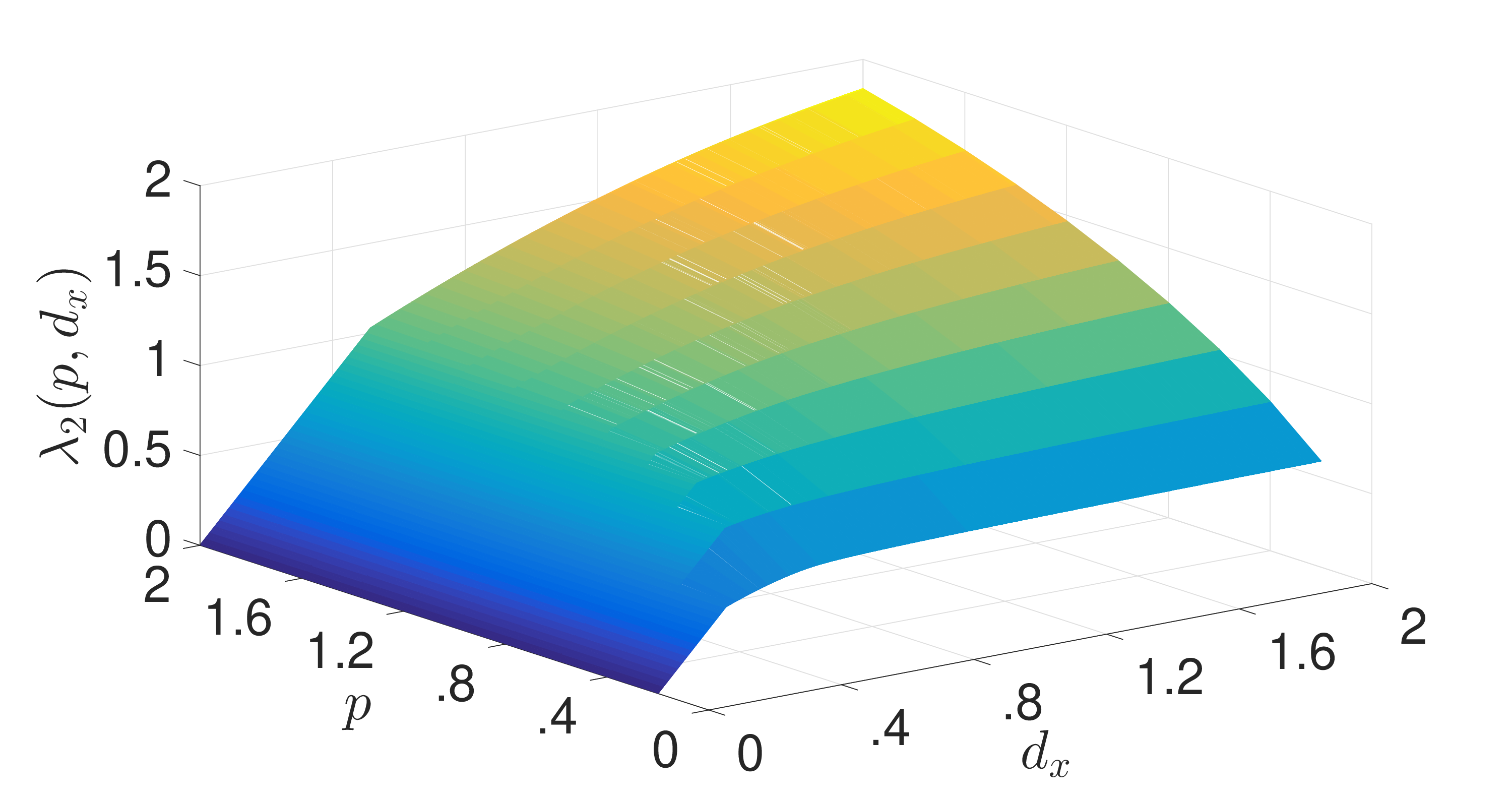} 
\caption{Plot of Algebraic connectivity $\lambda_2 (d_x,p)$ \textbf{for multiplex network (unweighted) designed using the BA model} for the perturbed network layers, with $0.2 \leq p \leq 2.0$ and variation in inter-layer link weights $0.2 \leq d_x \leq 2.0$. Variation in $\lambda_2$ (different color shades) is observed for the given values of $p$ and $d_x$ implying that there is an effect of perturbation in the network layers of the multiplex network.}

\label{ch3_baa1}
\end{center}
\end{figure}

\begin{figure}[!h]
\begin{center}
\includegraphics[width=1.0\linewidth,height=2.in]{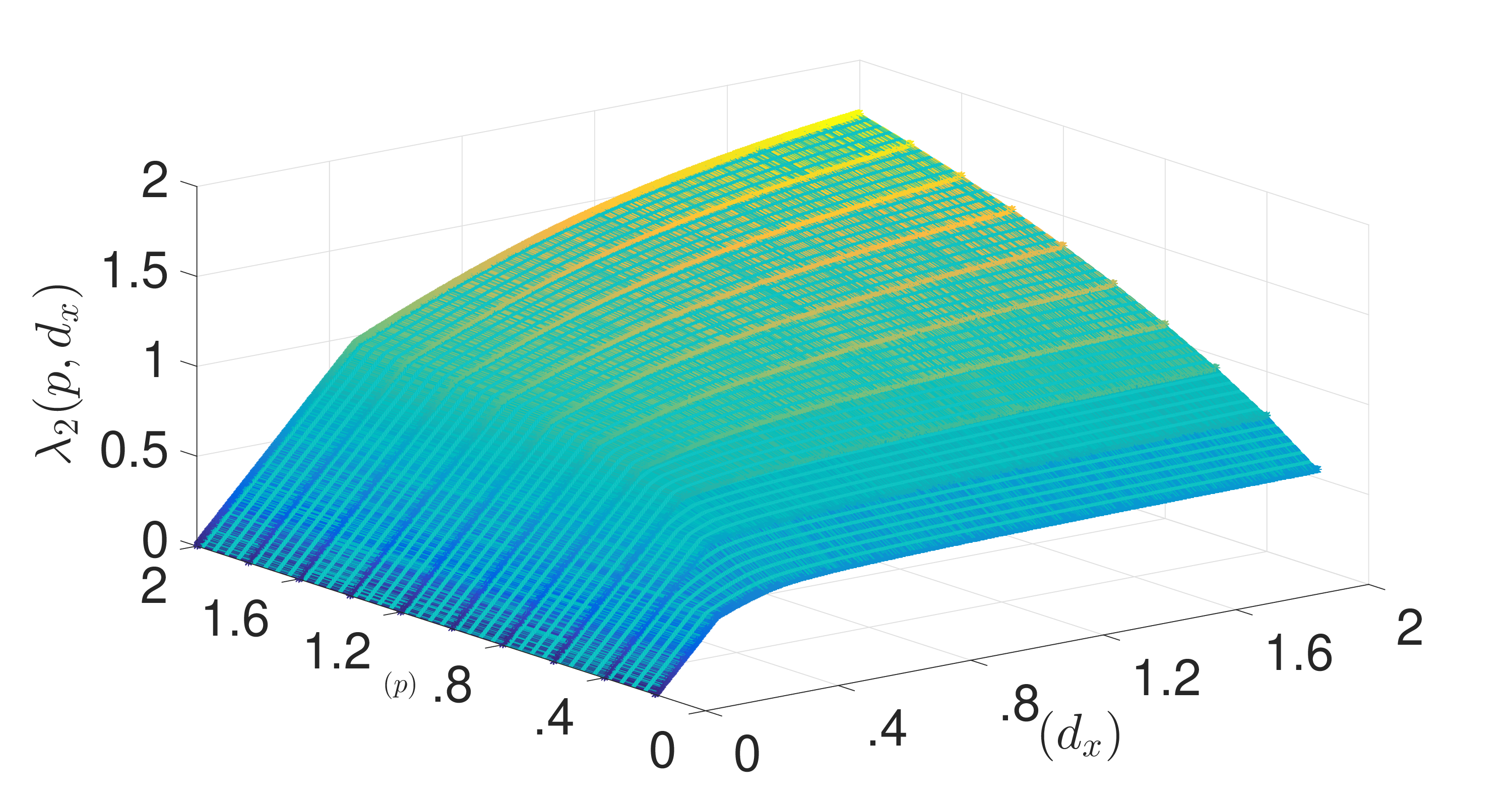} 
\caption{Plot of Algebraic connectivity $\lambda_2 (d_x,p)$ \textbf{for multiplex network (weighted) designed using the BA model} for the perturbed network layers, with $0.2 \leq p \leq 2.0$ and variation in inter-layer link weights $0.2 \leq d_x \leq 2.0$. Variation in $\lambda_2$ (different color shades) is observed for the given values of $p$ and $d_x$ implying that there is an effect of perturbation in the network layers of the multiplex network.}
\label{ch3_baa2}
\end{center}
\end{figure}

Simulation results for $\lambda_2 (d_x,p)$ in the unweighted case are plotted in Fig. \ref{ch3_baa1}. It is observed that for lower values of $(p=0.2)$, $\lambda_2$ grows monotonically (as a straight line) with increasing $d_x$, and $\lambda_2$ saturates at $d_x > 0.550$. However, when the network layers are perturbed, the effect of $d_x$ becomes significant, and $\lambda_2 (d_x > 0.550, p > 0.4)$ experiences an abrupt increase and does not stabilize even at $d_x=2$. In the regime where $d_x > 0.550$, $\lambda_2$ is controlled by a monotonically increasing function that stabilizes at $\lambda_2$ of ${(\mathcal{L}^1 + \mathcal{L}^2)/2}$, with ${\mathcal{L}^1 + \mathcal{L}^2}$ being the Laplacian matrix of the aggregated network, as shown in Fig. \ref{ch3_baa1}. For the weighted case, not much change is observed in the behavior of $\lambda_2 (d_x,p)$, except at $d_x \approx 0.450$, where $\lambda_2 (d_x,p)$ shows an abrupt change, as shown in Fig. \ref{ch3_baa2}.

\subsubsection{Network Layers of Multiplex Network Designed Using Power Law Network Model}\label{ch3_result_algebraic_Config}
\begin{figure}[!h]
\begin{center}
\includegraphics[width=1.0\linewidth,height=2.in]{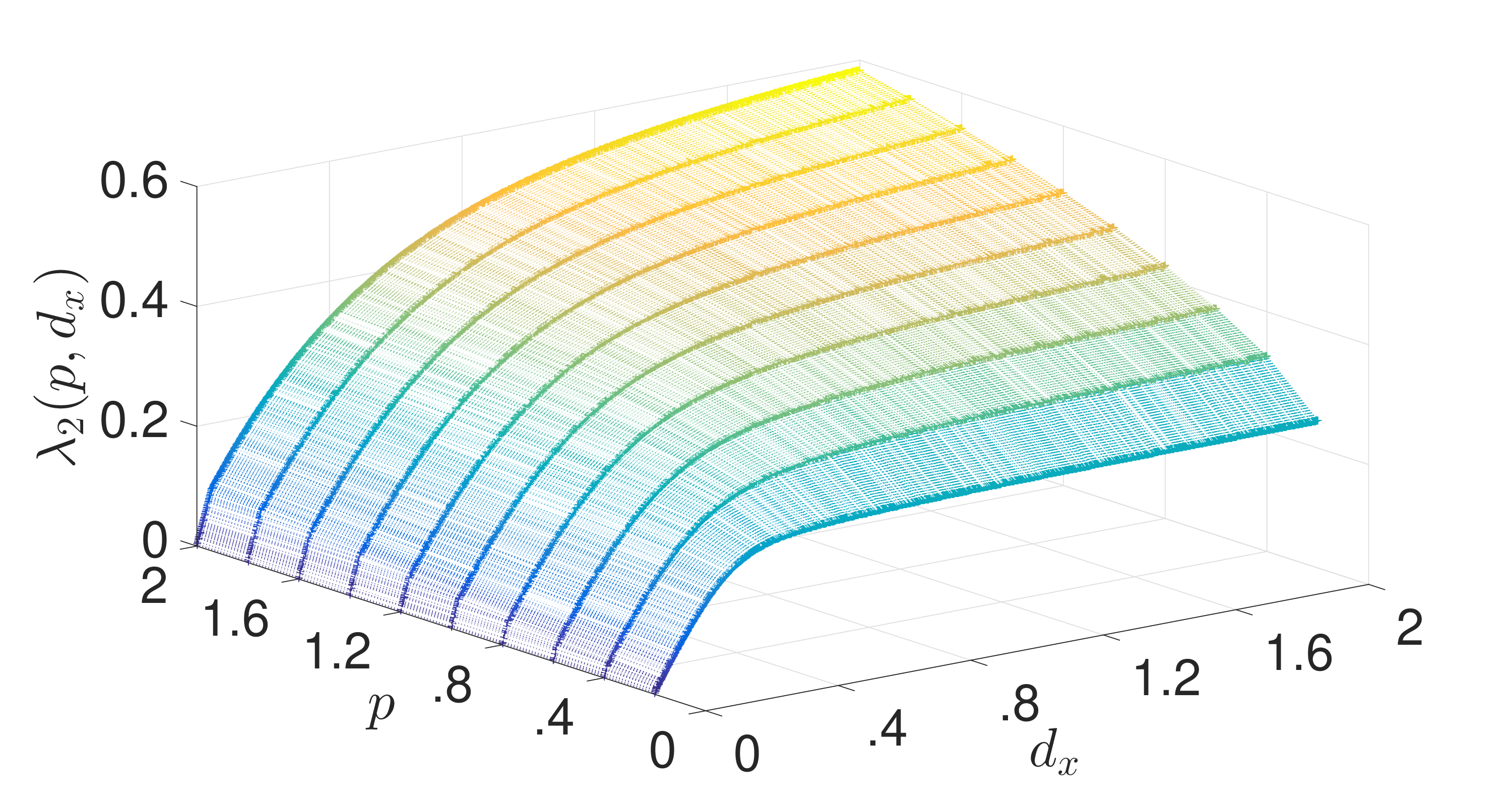} 
\caption{Plot of Algebraic connectivity $\lambda_2 (d_x,p)$ \textbf{for multiplex network (unweighted) designed using the Power law network model} for the perturbed network layers, with $0.2 \leq p \leq 2.0$ and variation in inter-layer link weights $0.2 \leq d_x \leq 2.0$. Variation in $\lambda_2$ (different color shades) is observed for the given values of $p$ and $d_x$ implying that there is an effect of perturbation in the network layers of the multiplex network. }

\label{ch3_config1}
\end{center}
\end{figure}

For the unweighted case, it is evident from the simulations results that similar to the BA model case, $\lambda_2$ grows monotonically and gets saturated at ($d_x > 0.8,p \approx 0.2$) as depicted in Fig. \ref{ch3_config1}.  Introducing more perturbation to the network layers results in a sudden change in $\lambda_2$ at lower values of $d_x$ (Fig. \ref{ch3_config1} shows an abrupt change at $\lambda_2 (d_x \approx 0.1, p \approx 2)$). Additionally, variations in the values of ($d_x, p$) compared to the BA model result in deviations in $\lambda_2$. For the BA model, $\lambda_2 (d_x=2, p=2) \approx 1.8$, while for the Power law network, $\lambda_2 (d_x=2, p=2) \approx 0.6$. This difference arises due to the distinct topological structures, specifically the higher average clustering coefficients observed in Table \ref{ch3_cc1}. In the weighted case, $\lambda_2$ grows and, after a quick change at ($d_x > 0.5, p \approx 0.2$), stabilizes, as shown in Fig \ref{ch3_config1}. The effect of perturbation (increasing $p$ in the network layers) on $\lambda_2$ is similar to the unweighted case, with a slight shift in $\lambda_2 (d_x, p)$. However, unlike the uniform weight case, $\lambda_2 (d_x=2, p=2) \approx 0.38$.

Indeed, in Small World (SW) and Scale-Free (SF) networks, large values of clustering produce a hindrance to global synchronization since the network partitions into clusters that oscillates at different frequencies\cite{mcgraw2005}.

\begin{figure}[!h]
\begin{center}
\includegraphics[width=1.0\linewidth,height=2.in]{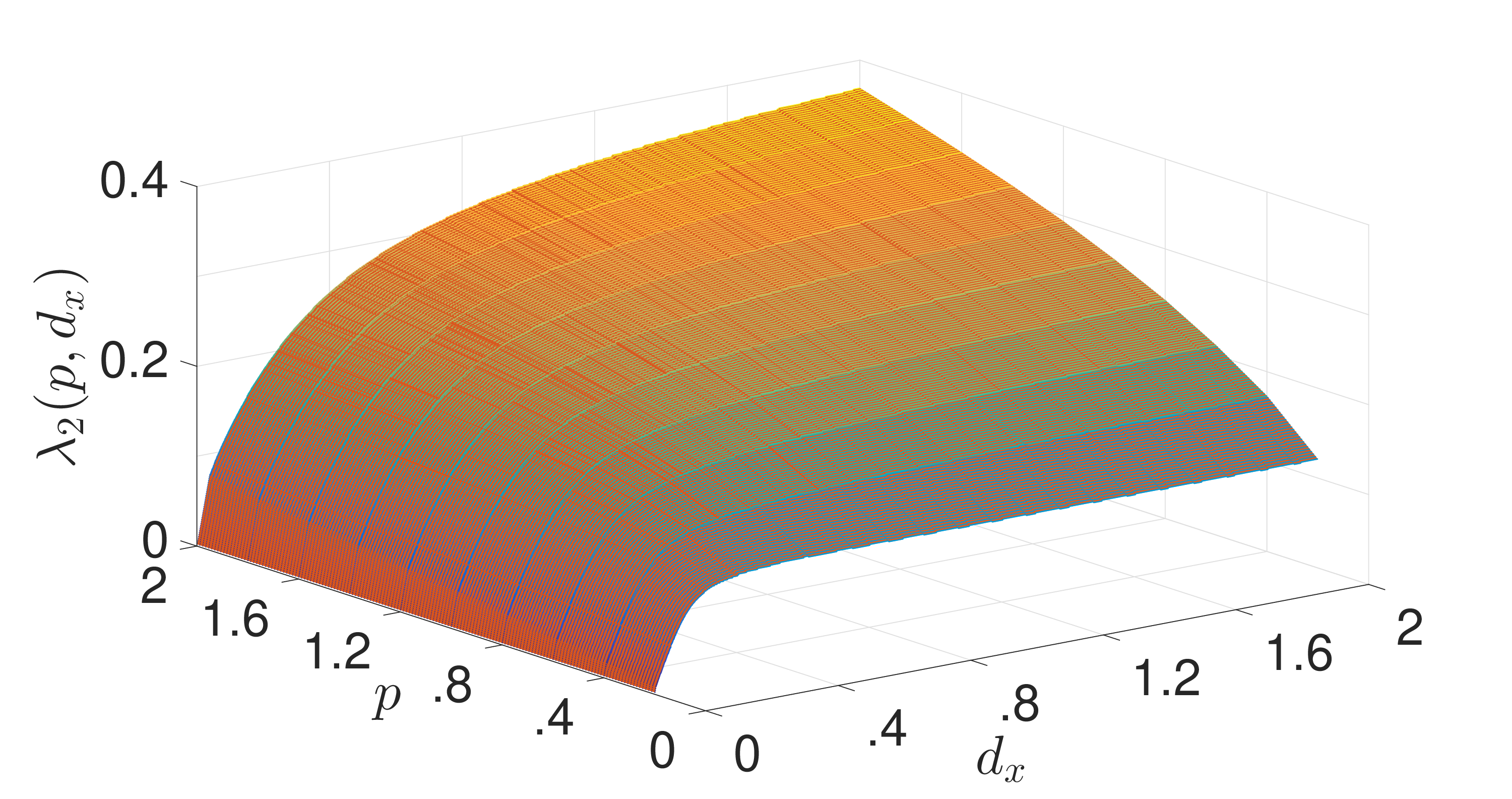} 
\caption{Plot of Algebraic connectivity $\lambda_2 (d_x,p)$ \textbf{for multiplex network (weighted) designed using the Power law network model} for the perturbed network layers, with $0.2 \leq p \leq 2.0$ and variation in inter-layer link weights $0.2 \leq d_x \leq 2.0$. Variation in $\lambda_2$ (different color shades) is observed for the given values of $p$ and $d_x$ implying that there is an effect of perturbation in the network layers of the multiplex network.}
 
\label{ch3_config2}
\end{center}
\end{figure}

\subsubsection{Multiplex Network Constructed From CS-Aarhus Social Multiplex Network \cite{magnani2013} Dataset} \label{ch3_result_algebraic_CS}
\begin{figure}[!h]
\begin{center}
\includegraphics[width=1.0\linewidth,height=2in]{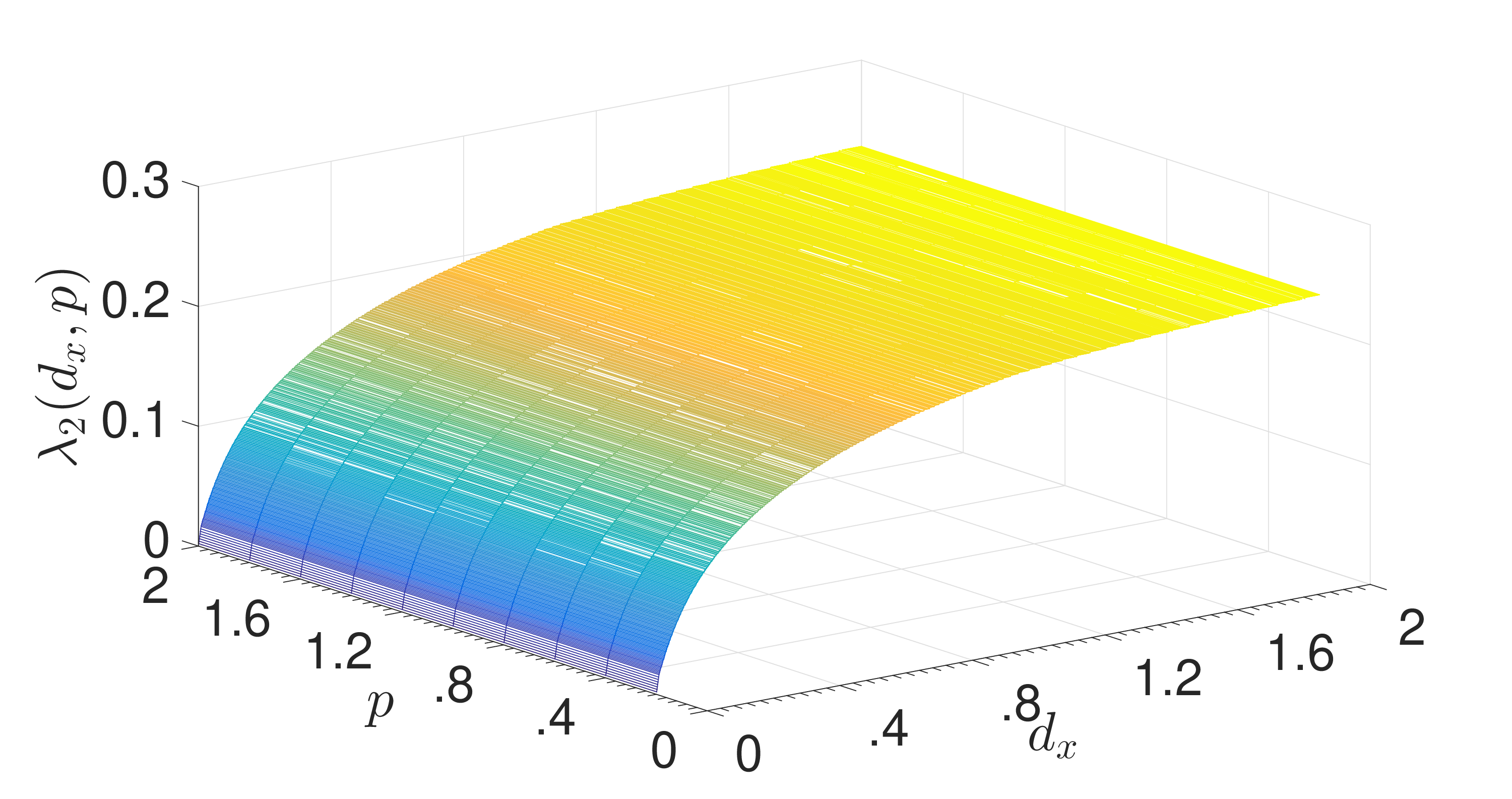} 
\caption{Plot of Algebraic connectivity $\lambda_2 (d_x,p)$ \textbf{for multiplex network (unweighted) constructed from CS-Aarhus Social Multiplex Network \cite{magnani2013} Dataset} for the perturbed network layers, with $0.2 \leq p \leq 2.0$ and variation in inter-layer link weights $0.2 \leq d_x \leq 2.0$. The apperance of uniform colors shades (Yellow, Saffron, Green, Blue) indicates that there is not much effect of perturbation in the network layers and $\lambda_2$ receives nearly same values for given values of $d_x$ and $p$.
 }

\label{ch3_real1}
\end{center}
\end{figure}

\begin{figure}[!h]
\begin{center}
\includegraphics[width=1.0\linewidth,height=2in]{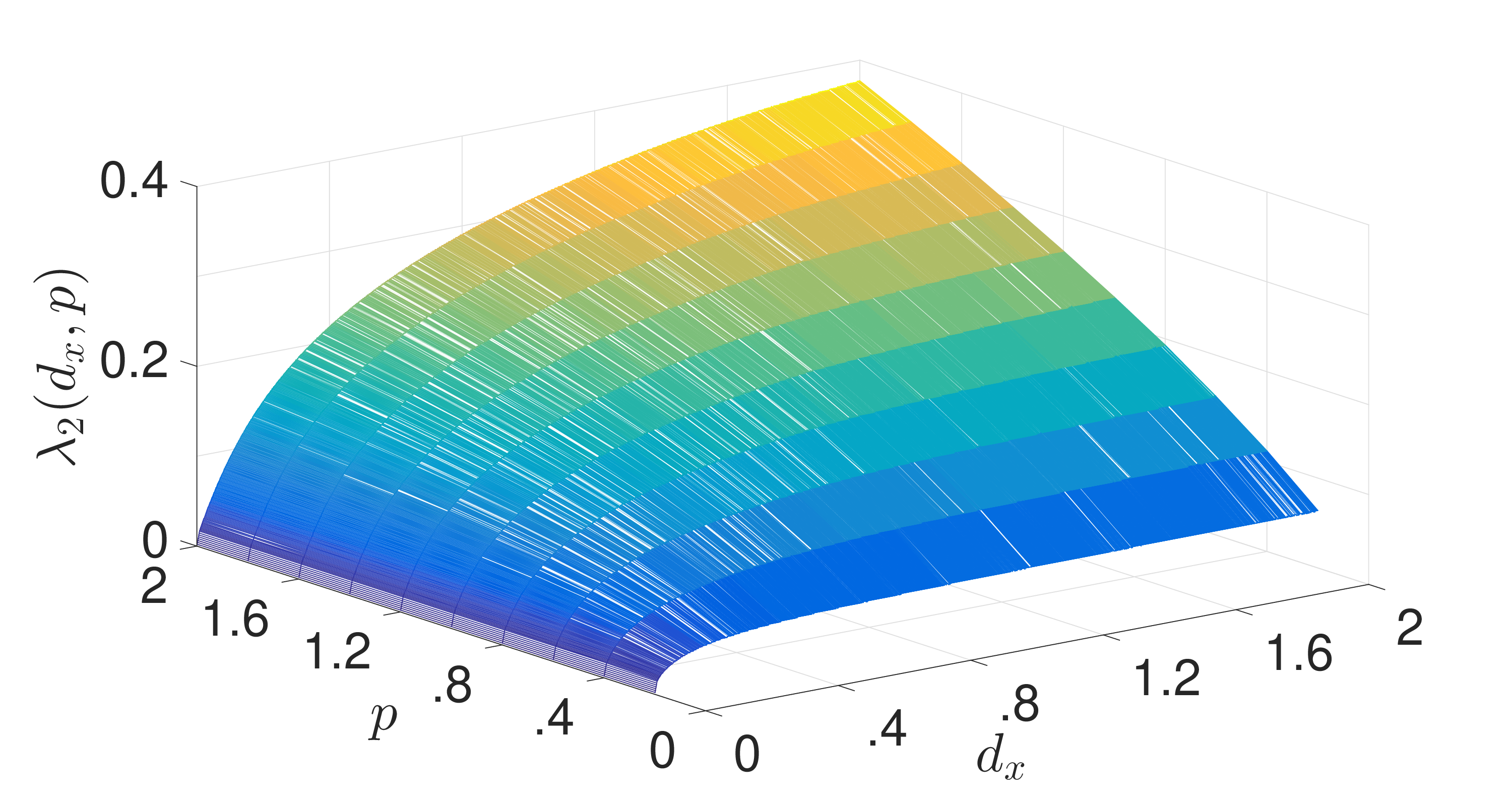} 

\caption{Plot of Algebraic connectivity $\lambda_2 (d_x,p)$ \textbf{for multiplex network (weighted) constructed from CS-Aarhus Social Multiplex Network \cite{magnani2013} Dataset} for the perturbed network layers, with $0.2 \leq p \leq 2.0$ and variation in inter-layer link weights $0.2 \leq d_x \leq 2.0$. Variation in $\lambda_2$ (different color shades) is observed for the given values of $p$ and $d_x$ implying that there is an effect of perturbation in the network layers of the multiplex network.}
 
 \label{ch3_real2}
\end{center}
\end{figure}

In the unweighted case, $\lambda_2$ shows an abrupt change at lower values of $d_x \approx 0.1$ for all considered values of $p$, and it stabilizes at $d_x \approx 0.8$, achieving a maximum value of $\lambda_2 \approx 0.24$, as shown in Fig. \ref{ch3_real1}. The appearance of uniform colors (Yellow, Saffron, Green, Blue) indicates that perturbations in the network layers have little effect, and $\lambda_2$ attains nearly the same values for given values of $d_x$ and $p$, as represented in Fig. \ref{ch3_real1}. However, in the weighted case, $\lambda_2$ shows an abrupt change at $d_x \approx 0.1$ and stabilizes at $\lambda_2 \approx 1.8$ for $d_x > 0.1$ at $p=0.2$. Unlike the unweighted case, variations in $\lambda_2$ (different color shades in Fig. \ref{ch3_real2}) are observed for the given values of $p$ and $d_x$, implying that perturbations affect the network layers of the multiplex network, as represented in Fig. \ref{ch3_real2}.

Simulation results obtained from synthetic and empirical multiplex network datasets reveal that the topological structure ($\langle CC \rangle$), intra-layer perturbations in link weights, and inter-layer link weights all affect the algebraic connectivity.

\subsection{ Stability of Synchronization} \label{ch3_result_stability}

In this section, we analyze the effect of inter-layer edge weights (controlled by the parameter $d_x$) and the average clustering coefficient ($\langle CC \rangle$) on the stability of the synchronization process ($R$) for the considered weighted and unweighted multiplex networks. A lower value of $R$ indicates maximum stability \cite{barahona2002}. The optimal value of $R$ for the multiplex network is approximated by $d_x$ at which analytical curves plotted using Equations \eqref{ch3_weak} and \eqref{ch3_strong} intersect each other.

\begin{figure}[!h]
\begin{center}
\includegraphics[width=0.9\linewidth,height=2in]{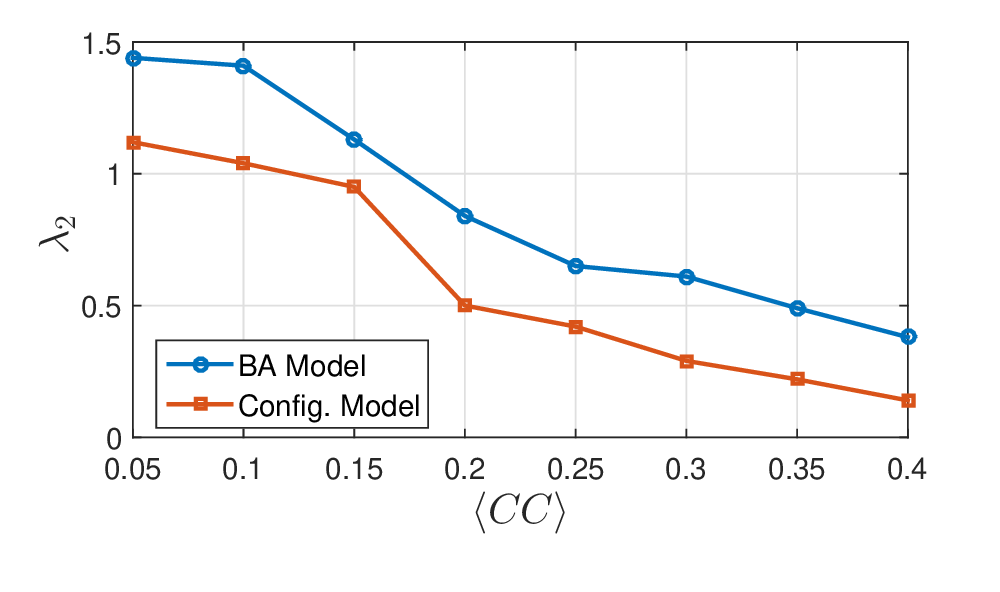} 
\caption{Variation of $\lambda_2$ against average clustering coefficient $\langle CC \rangle$.}
 
 \label{ch3_cc}
\end{center}
\end{figure}

When two independent network layers are interconnected to form a multiplex network, the values of $\langle CC \rangle$ differ from those in the individual layers. This occurs due to the addition of new edges between nodes in one layer and their replicas in the other layer. These changes are shown in Table \ref{ch3_cc1} for single-layer networks and multiplex networks. Additionally, there is a variation of $\lambda_2$ against $\langle CC \rangle$. An increase in the value of $\langle CC \rangle$ leads to a decline in the value of $\lambda_2$, as represented in Fig. \ref{ch3_cc}. This occurs because the network splits into dynamical clusters, with nodes within each cluster being more connected compared to the connections among different clusters.

\begin{table}[h]
\caption{Values of average clustering coefficient $(\langle CC \rangle)^L$ (the individual layers), $(\langle CC \rangle)^M$, $\lambda_2^M$, $\lambda_{N}^M$ and $k^M_{max}$ of multiplex networks constructed using BA model and Empirical data-set of CS-Aarhus.}
\begin{center}
\begin{tabular}{ |p{3.2cm}|p{1.2cm}|p{1.2cm}|p{1.2cm}|p{.9cm}|p{.9cm}|p{.9cm}|p{.9cm}|p{.9cm}|p{.9cm}|  }
 
 \hline
 \textbf{Parameters} $\Longrightarrow$ & ${\langle CC \rangle}^{L_1}$ & ${\langle CC \rangle}^{L_2}$ & ${\langle CC \rangle}^{M}$  & $\lambda_2^M$ & $\lambda_{N}^M$ & $k_{max}$ \\
 \hline
 BA model  & 0.079 & 0.085 & 0.05 & 1.42 & 48.37 &50\\
 \hline
 Power Law model & 0.17 & 0.13 & 0.07 & 0.48 &120.8 & 122\\
 \hline
 CSA    &0.67 & 0.63 & 0.41 & 0.76 & 28.14 &28\\
 
 \hline
\end{tabular}
\end{center}
\label{ch3_cc1}
\end{table}

\subsubsection{Network Layers of Multiplex Network Designed Using BA Model} \label{ch3_result_stability_BA}

\begin{figure}[!htb]
\begin{center}
$\begin{array}{ccc}
\includegraphics[width=.5\linewidth,height=2in]{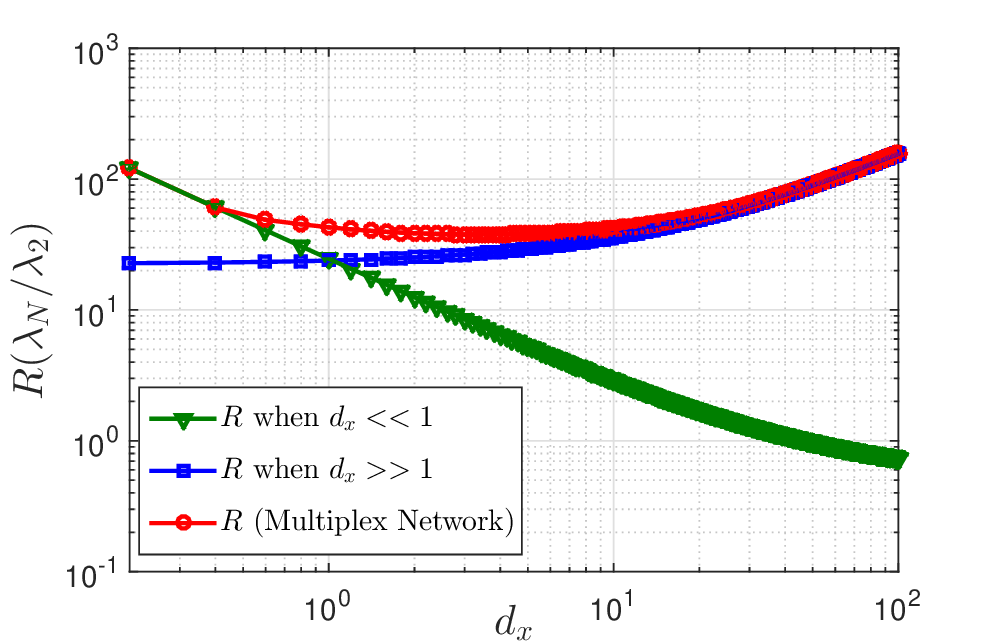}  &
\includegraphics[width=.5\linewidth,height=2in]{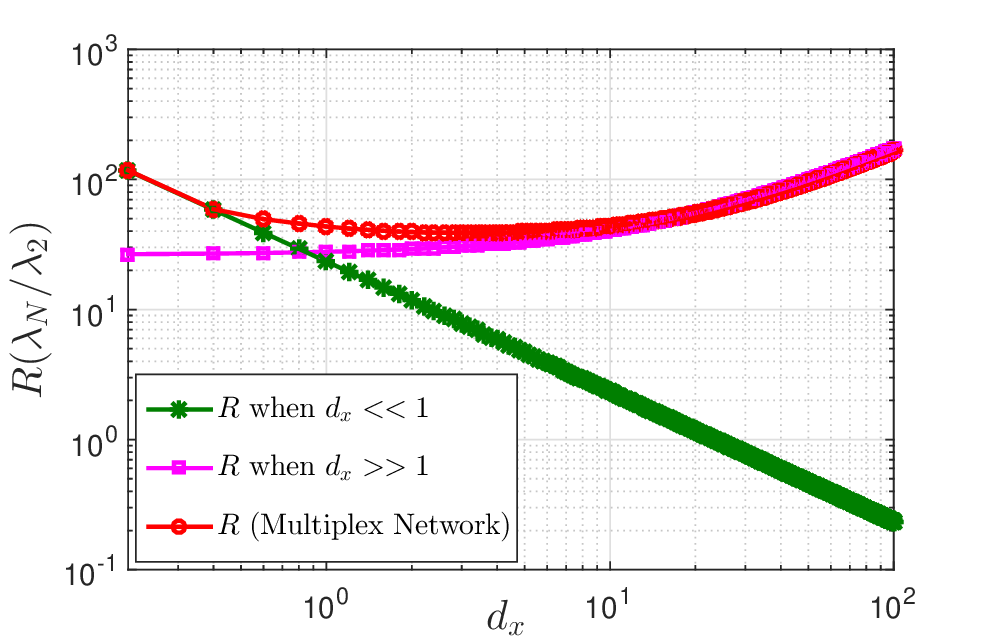} \\
\mbox{(a)} & \mbox{(b)} \\
\end{array}$
\caption{Eigenratio ($R$) against $d_x$ for BA model (a) for unweighted case (b) for weighted case. Optimal value of $R$ is approximated by point of intersection of two analytical curves obtained from Eqs. \ref{ch3_weak} and \ref{ch3_strong}, respectively.}
\label{ch3_R_ba}
\end{center}
\end{figure}

The values of parameter $R$ obtained from simulation are plotted (Red color curve) against $d_x$ for the considered unweighted and weighted multiplex networks, as shown in Figs. \ref{ch3_R_ba} (a) and (b), respectively. Theoretical values of $R$ approximated by Equations \eqref{ch3_weak} and \eqref{ch3_strong} are represented by green and blue curves, respectively, in the same figure. As $d_x$ increases, the values of $R$ obtained using Equations \eqref{ch3_weak} and \eqref{ch3_strong} decrease and increase continuously, respectively.  It is observed from Figs. \ref{ch3_R_ba} (a) and (b) that the optimal value of $R \approx 45$ is obtained at $d_x=1$ for the unweighted case, and $R \approx 49$ is achieved at $d_x=0.7$ for the weighted case. 

Interestingly,  This behavior is due to the properties of $\lambda_2$ as demonstrated in Figs. \ref{ch3_baa1} and \ref{ch3_baa2}. For the initial (lower) values of $d_x$, both $\lambda_2$ and $\lambda_2'$ increase, which decreases $R$. However, at higher values of $d_x$ and lower values of $p=0.2$, $\lambda_2$ stabilizes (as shown in Figs. \ref{ch3_baa1} and \ref{ch3_baa2}) while $\lambda_{N}$ of the supra-Laplacian matrix continues to increase. Thus, we find that weighted and unweighted multiplex networks differ in terms of $d_x$ and $R$. An interesting point to note that the values of $R$ against $d_x$ first decrease and, after reaching a minimum (optimal) value, begin to increase. This behavior is due to the properties of $\lambda_2$ as demonstrated in Figs. \ref{ch3_baa1} and \ref{ch3_baa2}. For the initial (lower) values of $d_x$, both $\lambda_2$ and $\lambda_2'$ increase, which decreases $R$. However, at higher values of $d_x$ and lower values of $p=0.2$, $\lambda_2$ stabilizes (as shown in Figs. \ref{ch3_baa1} and \ref{ch3_baa2}) while $\lambda_{N}$ of the supra-Laplacian matrix continues to increase. Thus, we find that weighted and unweighted multiplex networks differ in terms of $d_x$ and $R$.

\begin{table}[htb!]
\begin{center}
\caption{ Represents the optimum values of controlling parameters $d_x$ for which eigenratio ($R$) is minimum for the BA model and Empirical dataset. }
\begin{tabular}{ |p{3.2cm}|p{2cm}|p{2cm}|p{1.5cm}|p{1.5cm}|}
 
 \hline
 \textbf{Parameters} $\Longrightarrow$ & $d_x (Unwtd)$ & $R (Unwtd)$ & $d_x (Wtd)$ & $R  (Wtd)$ \\
 \hline
 BA model  & 1 & 44.48 & 0.7 & 48.67\\
 \hline
 
 Power Law model & 0.5 & 177 & 0.38 & 235.61\\
 \hline
 CSA    &0.75 & 24.98 & 0.25 & 66.25 \\
 
 \hline
\end{tabular}
\label{ch3_optimal_value}
\end{center}
\end{table}


\subsubsection{Network Layers of Multiplex Network Designed Using Power Law Network Model}
\label{ch3_result_stability_Config}
\begin{figure}[!htb]
\begin{center}
$\begin{array}{ccc}
\includegraphics[width=.5\linewidth,height=2in]{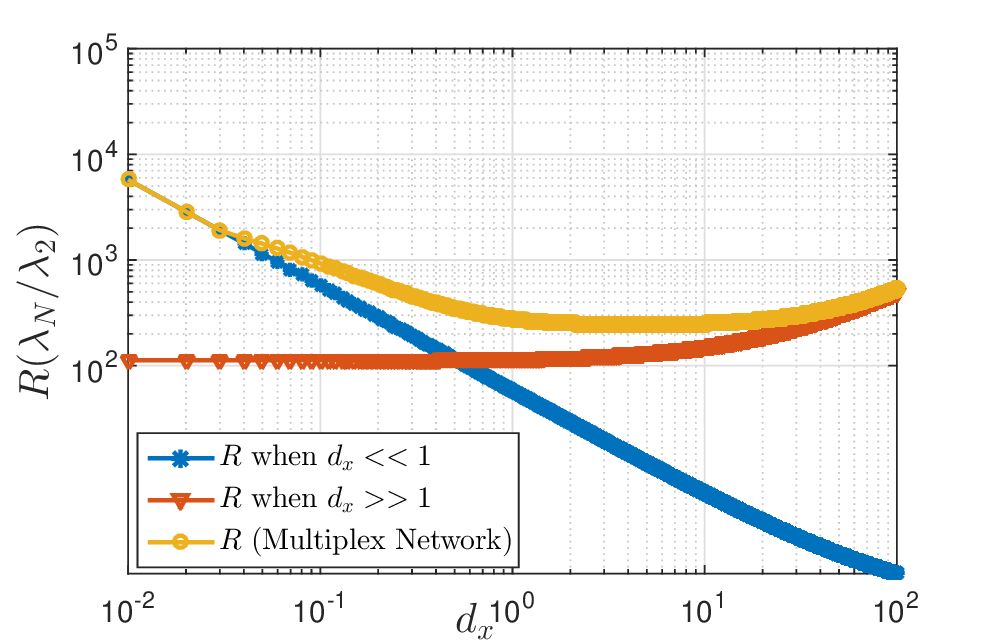}  &
\includegraphics[width=.5\linewidth,height=2in]{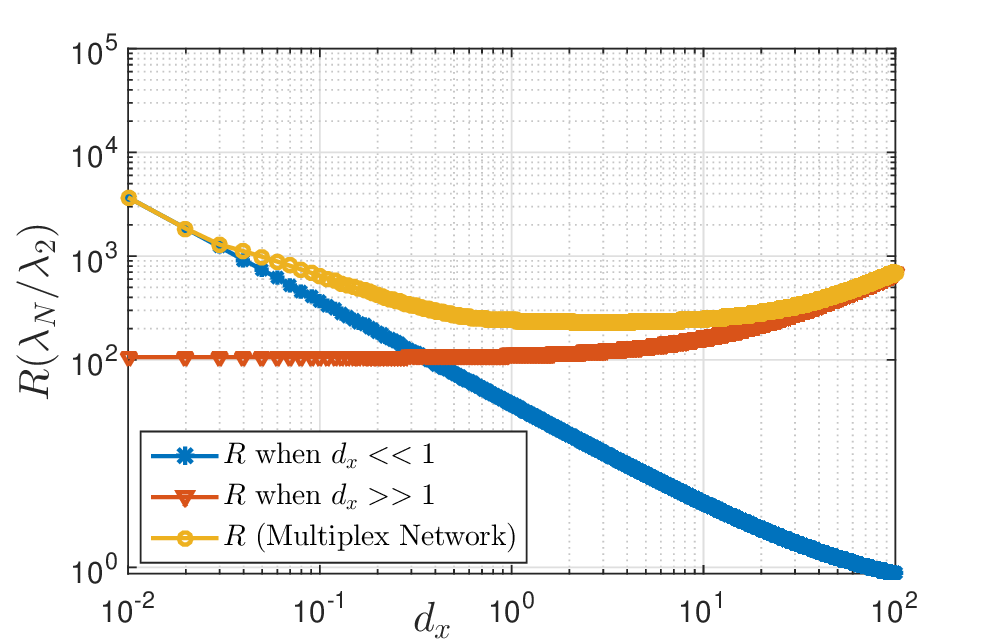} \\
\mbox{(a)} & \mbox{(b)} 
\end{array}$
\caption{Eigenratio ($R$) against $d_x$ for Power Law Network model (a) for unweighted case (b) for weighted case. Optimal value of $R$ is approximated by point of intersection of two analytical curves obtained from Eqs. \ref{ch3_weak} and \ref{ch3_strong} respectively.}
\label{ch3_R_config}
\end{center}
\end{figure}

In this case, optimal value of $R \approx 177$ is obtained at $d_x \approx 0.5$ for unweighted cases as well as $R\approx 235$ is achieved at $d_x \approx 0.38$ for weighted case as shown in the Figs. \ref{ch3_R_config} (a) and (b) respectively. It occurs due to a difference in the topological characteristics as compared to the multiplex network using the BA model. It is observed from Figs. \ref{ch3_config1} and \ref{ch3_config2} that for the given values of $d_x$ and $p$, values of $\lambda_2$ is  lower in the case of the Power Law Network model as compared to the BA model due to the effect of average clustering coefficient ($\langle CC \rangle$). At the same time, from Table \ref{ch3_cc1} we find that $\lambda_{N}$ is higher (approx. three times) in the case of multiplex network constructed using Power Law Network model as compared to BA model. Thus, the difference in $\lambda_{N}$ and $\lambda_2$ produces variation in the parameter $R$.

\subsubsection{Multiplex Network Constructed From CS-Aarhus Social Multiplex Network \cite{magnani2013} Dataset} \label{ch3_result_stability_CS}
\begin{figure}[!htb]
\begin{center}
$\begin{array}{ccc}
\includegraphics[width=.5\linewidth,height=2in]{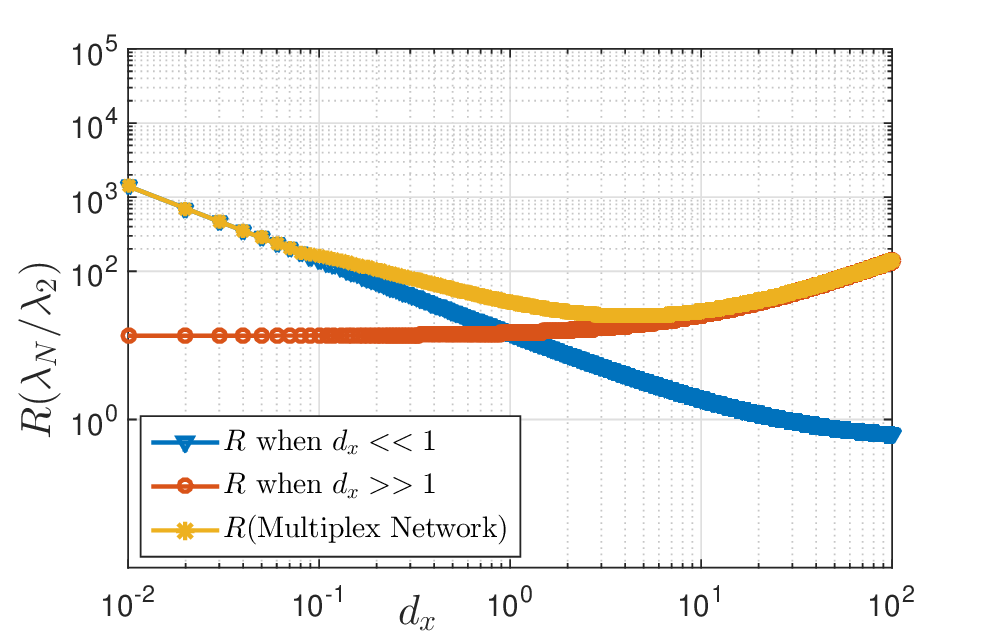}  &
\includegraphics[width=.5\linewidth,height=2in]{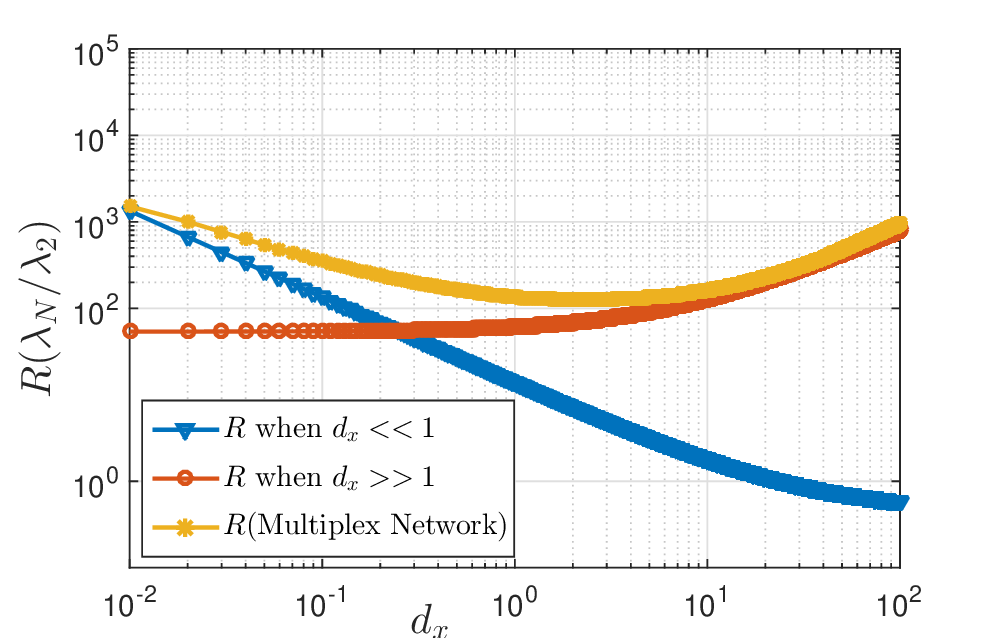} \\
\mbox{(a)} & \mbox{(b)} 
\end{array}$
\caption{Represent the plot of eigenratio ($R$) against $d_x$ for the multiplex network designed from CS-Aarhus social multiplex network \cite{magnani2013} dataset (a) unweighted case (b) weighted case. Optimal value of $R$ is approximated by point of intersection of two analytical curves obtained from Eq. \ref{ch3_weak} and \ref{ch3_strong} respectively.}
\label{ch3_R_real}
\end{center}
\end{figure}

It is observed from Figs. \ref{ch3_R_ba}, \ref{ch3_R_config} and \ref{ch3_R_real} that behaviour of $R$ is approx. same for the synthetic as well as empirical dataset multiplex networks. However, some variation in optimal values $R$ and $d_x$ is observed for weighted as well as unweighted cases. In this case, optimal values of $R \approx 25$ is obtained at $d_x \approx 0.75$ for unweighted case and $R \approx 66$ is achieved at $d_x \approx 0.25$ as shown in Fig. \ref{ch3_R_real}. Here, again the variation in $R$ and $d_x$ is noticed as compared to a multiplex network constructed using the BA model and Power Law Network model. The reason is the variation ib the topological characteristics of the network layers of the multiplex network.

Simulation results reveal that the stability of the synchronization $R$ is lower in the case of the Power Law Network model as compared to the BA model. However, multiplex networks constructed from the empirical dataset show higher stability $R$ than considered artificial multiplex networks. This variation is due to the topological characteristics (size, sparsity of intra-layer and inter-layer edges, etc.) of the network layers constituting multiplex networks. It is also observed from simulation results that there is variation in $R$ in the cases of the considered unweighted and weighted multiplex network.


\subsection{Time of Synchronization} \label{ch3_result_sync_time}
Synchronization time $(\tau)$ is when all the layers of the multiplex networks are fully synchronized $i.e$ when level of synchronizations $S\rightarrow 1$ at sufficiently large $\tau$. For the considered multiplex networks, values of the parameter $R$ are plotted against time $\tau$ for $d_x=0.3$ and $d_x=20$. For the simulation purpose, these values of $d_x$ are chosen to quantify lower and higher inter-layer link weights. However, any value of $d_x$ can be chosen.

\subsubsection{Network Layers of Multiplex Network Designed Using BA Model}\label{ch3_result_sync_time_BA}
\begin{figure}[!htb]
\begin{center}
\includegraphics[width=.8\linewidth,height=2.2in]{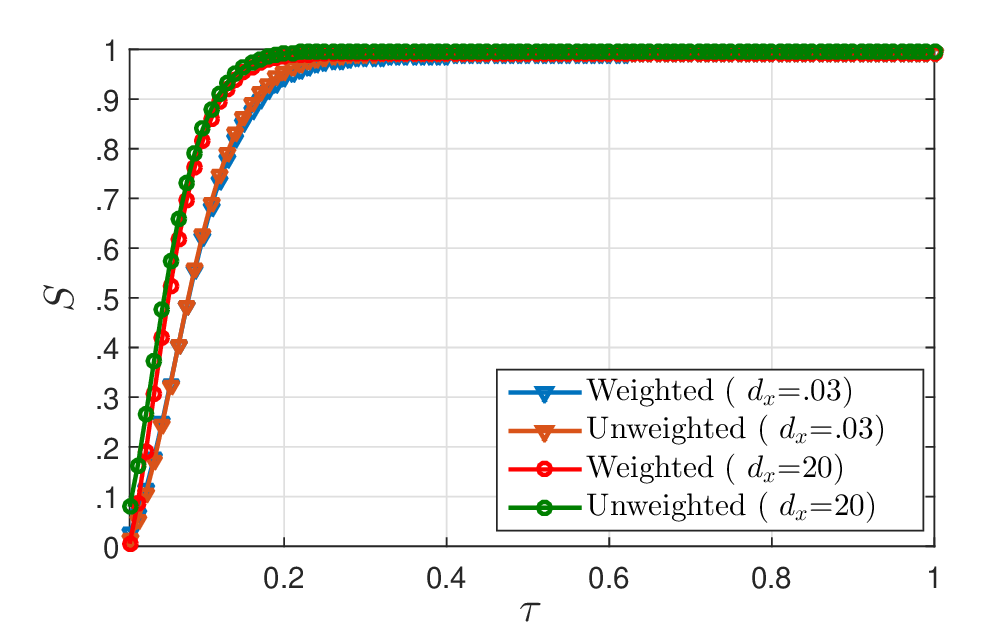}
\caption{Level of synchronization ($S$) as a function of $\tau$ for multiplex network using \textbf{BA model}. Values of $S$ have been plotted for weighted and unweighted cases with $d_x = 0.03$ and $d_x = 20$. The parameter $S\rightarrow 1$ at earlier timestamps $\tau$ in  unweighted and weighted cases with $d_x = 20$ as compared to $d_x=0.3$ in both the multiplex networks.}
\label{ch3_time_BA}
\end{center}
\end{figure}

The parameter $S$ obtained from Equation \eqref{ch3_time} is plotted against time-scale  $\tau$ as shown in Fig. \ref{ch3_time_BA}. For the lower values of $d_x=0.3$, the values of $S$ against $\tau$ are nearly same for weighted as well as unweighted cases. At higher $d_x=20$, the parameter $S \rightarrow 1$ at $\tau\simeq 0.2$ whereas for $d_x=0.3$, $S \rightarrow 1$ at $\tau\simeq 0.4$ as demonstrated in Fig. \ref{ch3_time_BA}. Higher values of $d_x$ is an indication of increasing the frequency of inter-layer interaction between the nodes present at the different network layers of the considered multiplex network. Thus, we find that increasing the value of $d_x$ lead to establish the synchronization in multiplex networks more quickly.

\subsubsection{Network Layers of Multiplex Network Designed Using Power Law Network Model}
\label{ch3_result_sync_time_Config}
\begin{figure}[!htb]
\begin{center}
\includegraphics[width=.8\linewidth,height=2.2in]{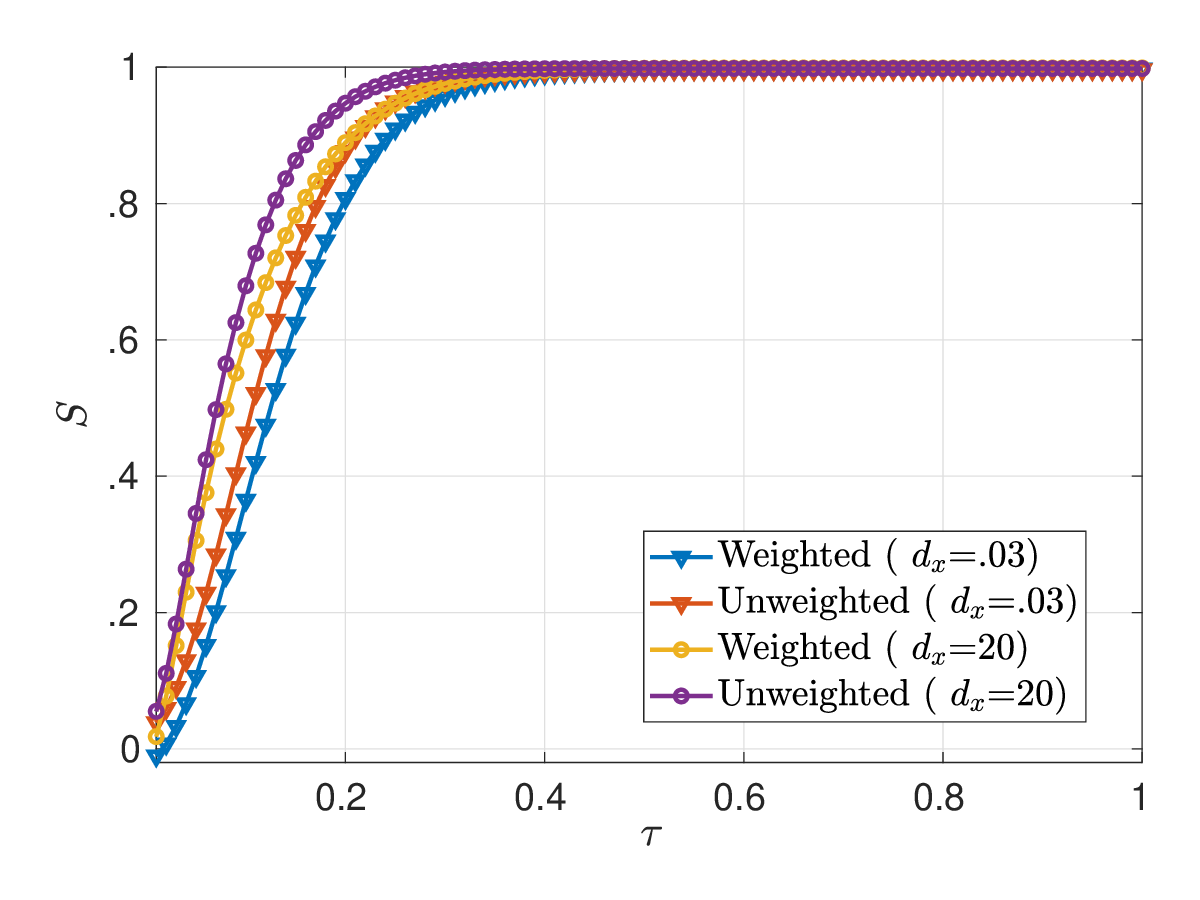} 
\caption{Level of synchronization ($S$) as a function of $\tau$ for multiplex network using \textbf{Power Law Network model}. The parameter $S\rightarrow 1$ for $d_x=20$ at $\tau \approx 0.3, 0.4$ for unweighted and weighted cases respectively. At $\tau=2$, $S \approx 0.8,0.9$ for weighted and unweighted cases for $d_x=0.3$. }
\label{ch3_time_config}
\end{center}
\end{figure}

In this case, for $d_x=0.3$ and $d_x=20$ the parameter $S$ lags in approaching 1 for weighted case as compared to unweighted cases. It occurs due the fact that in weighted case, non-uniform link weights (intra-layer and inter-layer) cause change in the state (information) of the nodes in the multiplex such that $S\rightarrow 1$ at later time $\tau$ as compared to unweighted case. For $d_x=0.3$, $S\rightarrow 1$ at $\tau \approx 0.4$ for weighted as well unweighted cases. However, for $d_x=20$, $S\rightarrow 1$ at $\tau \approx 0.3$ and $\tau \approx 0.4$ for weighted and unweighted respectively as shown in Fig. \ref{ch3_time_config}.

\subsubsection{Multiplex Network Constructed From CS-Aarhus Social Multiplex Network \cite{magnani2013} dataset} \label{ch3_result_sync_time_CS}
\begin{figure}[!htb]
\begin{center}
\includegraphics[width=.8\linewidth,height=2.2in]{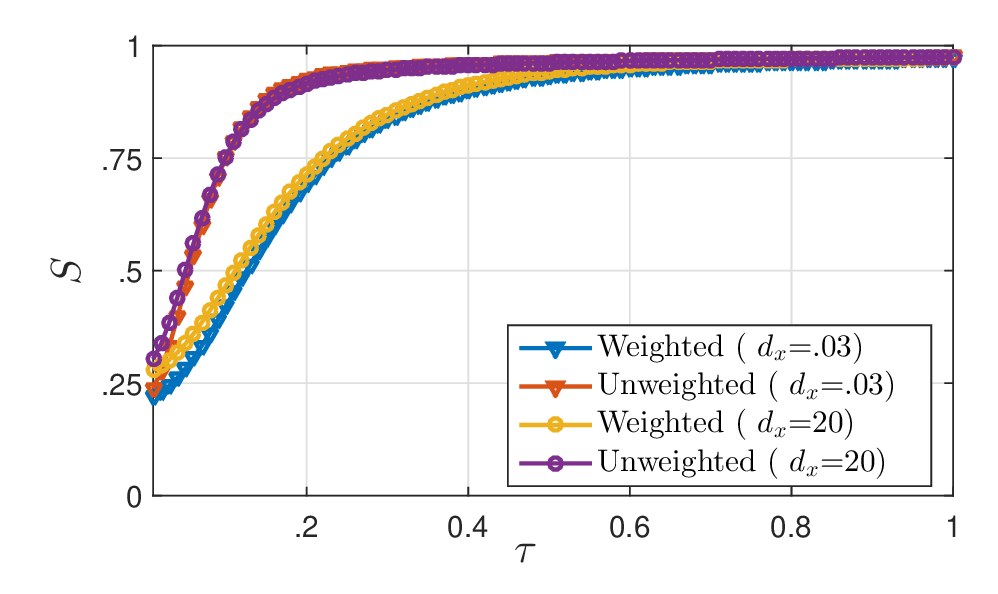}
\caption[Level of synchronization ($S$) as a function of $\tau$ for multiplex network using \textbf{Empirical data-set of online and offline relationship between the employees of the Computer Science department at Aarhus}. Values of $S$ have been plotted for weighted and unweighted cases with $d_x = 0.03$ and $d_x = 20$. The behavior of parameter $S$ is nearly same as in the case of multiplex constructed using BA model.]{Level of synchronization ($S$) as a function of $\tau$ for multiplex network using \textbf{ Empirical data-set of online and offline relationship between the employees of the Computer Science department at Aarhus}. Values of $S$ have been plotted for weighted and unweighted cases with $d_x = 0.03$ and $d_x = 20$. The behavior of parameter $S$ is nearly same as in the case of multiplex constructed using BA model. The value of $S\rightarrow 1$ at earlier times $\tau=0.3$ in unweighted and weighted cases with $d_x = 20$ as compared to $d_x=0.3$ in which $S\rightarrow 1$ at $\tau \approx 0.6$. }
\label{ch3_time_real}
\end{center}
\end{figure}

In this case, the behaviour of parameter $S$ against $\tau$ is nearly identical to the BA model. For $d_x=0.3$, $S \approx 0.7 $ and for $d_x=20$, $S \approx 0.7 $ at $\tau =2$ for weighted and unweighted cases as in Fig. \ref{ch3_time_real}. With the increase in time $\tau$, the gap between the values of $S$ keeps on decreasing and finally $S\rightarrow 1$ at $\tau=6$ for $d_x=0.3,20$. Thus, from Fig. \ref{ch3_time_real} we find that increasing the $d_x$ results in synchronization of the considered multiplex network at earlier time $\tau$.

\section{Conclusion and Future Work}
\label{conclusion_ch3}
In this paper, we examined the impact of perturbations in network layers on the algebraic connectivity ($\lambda_2$) and synchronization dynamics in unweighted and weighted multiplex networks, each comprising two network layers. Our simulation results revealed that for both synthetic and real dataset-based multiplex networks, $\lambda_2$ varies with the given value of $d_x$ and different values of $p$ across the considered individual multiplex networks and different topological multiplex networks. It occurs due the perturbation in the intra-layer link weights (individual multiplex networks) and the differences in the average clustering coefficient (${\langle CC \rangle}^{CSA} > {\langle CC \rangle}^{Power Law Network} > {\langle CC \rangle}^{BA}$). As far as stability of synchronization is concerned, multiplex networks constructed using the BA model show more stability than the Power Law Network model for weighted and unweighted cases. Maximum stability (minimum value of $R$) is obtained at $d_x=1,0.5$ for BA and Power Law Network models, respectively, for unweighted cases. Multiplex network constructed from CSA dataset shows the highest stability at $d_x=0.75$ for unweighted case. This variation instability occurs due to the characteristics of the network layer. In the case of a synthetic network, although the size (number of nodes) of network layers is the same, these differ in $\langle CC \rangle$, which produces variation in $\lambda_2$ and $\lambda_{N}$ of the supra-Laplacian matrix of the multiplex network thereby showing variation in the parameter $R$. 

Our simulation results show that multiplex networks are slightly less stable in the weighted case compared to the unweighted case due to non-identical intra-layer and inter-layer edge weights. For synchronization time $\tau$, the parameter $S \rightarrow 1$ is reached earlier when $d_x=20$ compared to $d_x=0.3$ in both weighted and unweighted cases. The time $\tau$ for $S \rightarrow 1$ is slightly higher in the weighted case for the Power Law Network model and CSA-dataset multiplex networks. This indicates that networks with higher average clustering coefficients ($\langle CC \rangle$) are less stable and take longer to achieve complete synchronization. However, tuning inter-layer edge weights can improve synchronization stability and reduce the time required. Therefore, the characteristics ($\langle CC \rangle$) of individual layers and inter-layer link weights significantly influence the synchronization process. In the future, this work can be extended to analyze graph energy in the synchronization process.

\section*{Acknowledgement}
We want to thank Dr Anurag Singh, Associate Professor at the Department for Computer Science and Engineering, National Institute of Technology Delhi, India for providing guidance to work in this domain during the doctorate work.

\bibliography{samplepaper}
\bibliographystyle{splncs04}

\end{document}